\documentclass[sigconf,nonacm]{acmart}
\usepackage{tikz}
\AtBeginDocument{%
  \providecommand\BibTeX{{%
    \normalfont B\kern-0.5em{\scshape i\kern-0.25em b}\kern-0.8em\TeX}}}

\makeatletter
\def\@ACM@checkaffil{% Only warnings
    \if@ACM@instpresent\else
    \ClassWarningNoLine{\@classname}{No institution present for an affiliation}%
    \fi
    \if@ACM@citypresent\else
    \ClassWarningNoLine{\@classname}{No city present for an affiliation}%
    \fi
    \if@ACM@countrypresent\else
        \ClassWarningNoLine{\@classname}{No country present for an affiliation}%
    \fi
}
\makeatother

\begin{document}

\title{S3C2 Summit 2023-06: \\ Government Secure Supply Chain Summit}

\author{William Enck$^{*}$, Yasemin Acar$^{\dagger}$, Michel Cukier$^{\ddagger}$, \\ Alexandros Kapravelos$^{*}$, Christian Kästner$^{\mathsection}$, Laurie Williams$^{*}$}

%% authors that will appear in the ACM reference format (without authormarks)
\def \authors{Mindy Tran, Yasemin Acar, Michel Cukier, William Enck, Alexandros Kapravelos, Christian Kästner, Laurie Williams}

\affiliation{%
    \institution{ $^*$North Carolina State University, Raleigh, NC, USA}
}
\affiliation{%
    \institution{$^\dagger$Paderborn University, Paderborn, Germany and George Washington University, DC, USA}
}
\affiliation{%
    \institution{$^\ddagger$University of Maryland, College Park, MD, USA}
}
\affiliation{%
    \institution{ $^\mathsection$Carnegie Mellon University, Pittsburgh, PA, USA}
}
%\email{{firstname.surname}@uni-paderborn.de}
%\email{mcukier@umd.edu}
%\email{{whenck, akaprav, lawilli3}@ncsu.edu}
%\email{kaestner@cs.cmu.edu}

\renewcommand{\shortauthors}{Secure Software Supply Chain Center (S3C2)}
\renewcommand{\shorttitle}{S3C2 Summit 2023-06: Government Secure Supply Chain Summit}

\begin{abstract}
  Recent years have shown increased cyber attacks targeting less secure elements in the software supply chain and causing fatal damage to businesses and organizations. Past well-known examples of software supply chain attacks are the SolarWinds or log4j incidents that have affected thousands of customers and businesses. The US government and industry are equally interested in enhancing software supply chain security. On June 7, 2023, researchers from the NSF-supported Secure Software Supply Chain Center (S3C2) conducted a Secure Software Supply Chain Summit with a diverse set of 17 practitioners from 13 government agencies.  
  The goal of the Summit was two-fold: 
  (1)~to share our observations from our previous two summits with industry, and 
  (2)~to enable sharing between individuals at the government agencies regarding practical experiences and challenges with software supply chain security.
  For each discussion topic, we presented our observations and take-aways from the industry summits to spur conversation.
  We specifically focused on 
  the Executive Order 14028, 
  software bill of materials (SBOMs), 
  choosing new dependencies,
  provenance and self-attestation, and
  large language models.
  The open discussions enabled mutual sharing and shed light on common challenges that government agencies see as impacting government and industry practitioners when securing their software supply chain. 
  In this paper, we provide a summary of the Summit. Full panel questions can be found at the beginning of each section and in the appendix.
\end{abstract}

\iffalse

\fi



%\received{30 September 2022}
%\received[revised]{1 December 2022}

\maketitle

\begin{tikzpicture}[overlay, remember picture]
\node[anchor=north west, %anchor is upper left corner of the graphic
      xshift=17.5cm, %shifting around
      yshift=-2.1cm] 
     at (current page.north west) %left upper corner of the page
     {\includegraphics[width=2.1cm]{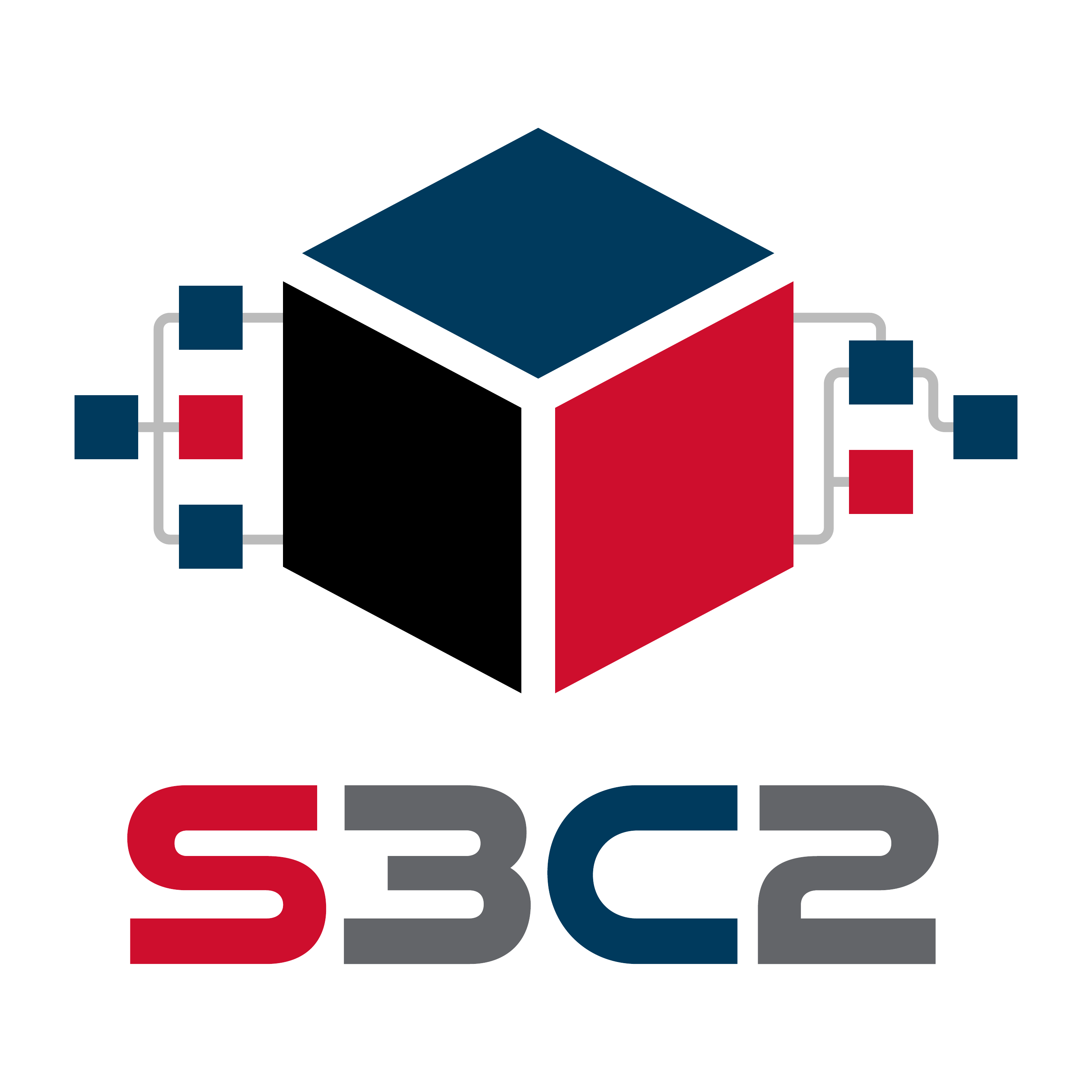}}; 
\end{tikzpicture}

\section{Introduction}

  Recent years have shown increased cyber attacks targeting less secure elements in the software supply chain and causing fatal damage to businesses and organizations. Past well-known examples of software supply chain attacks are the SolarWinds or log4j incidents that have affected thousands of customers and businesses. On June 7, 2023, two researchers from the NSF-supported Secure Software Supply Chain Center (S3C2) conducted a one-day Secure Software Supply Chain Summit with a diverse set of 17 practitioners from 13 government agencies.
  The goal of the Summit was two-fold: 
  (1)~to share our observations from our previous two summits with industry~\cite{summit1, Summit2}, and 
  (2)~to enable sharing between individuals at the government agencies regarding practical experiences and challenges with software supply chain security.

  Summit participants were recruited from 13 government agencies with interests in the security of software. Attendance was limited to keep the event small enough that honest communication between participants can flow. The Summit was conducted under the Chatham House Rules, which state that all participants are free to use the information discussed, but neither the identity nor the affiliation of the speaker(s), nor any other participant may be revealed. As such, none of the participating government agencies are identified in this paper.  

  The Summit consisted of five discussion topics led by one of the S3C2 researchers. Before the Summit, participants completed a survey to vote on the topics of the five discussion topics.  As such, the panel topics of interest to the government agencies. The remaining topics discussed in the industry summits were also briefly presented but with only minor discussion.  The questions posed to the panelists appear in the Appendix.   

  The two researchers (two professors) took notes on the discussions and created the first draft summary of the discussion based on these notes.  The draft was then by the other authors of this paper, who are also S3C2 researchers and experts in software supply chain security.

  The next seven sections provide a summary of the Secure Supply Chain Summit.

\section{Executive Order}
\label{sec:eo}

Executive Order 14028 \cite {EO} issued on May 12, 2021, charges organizations supplying critical software to the US government to improve the security and integrity of their software and the software supply chain.  Most organizations need to make procedural, operational, and cultural changes as a result. 

\subsection{What is the Goal?}

In contrast to the industry summits, the participants of the government summit are consumers of the EO requirements.
One participant noted that the EO does not apply to in-house developed software.
The participants reflected on what they see through their interactions with contractors.

The discussion began by reflecting on the ambiguity concerns raised in the industry summits~\cite{summit1, Summit2}.
One participant acknowledged that when it comes to operationalizing the EO, a lot is still in flight.
Some deadlines of the EO have been missed, and there is uncertainty about when things will go into effect.
Another participant gave valuable insight into the ambiguity problems raised by industry:
``Ambiguity from a government perspective also means choice for the company.''
The government is careful not to be too prescriptive about specific technologies, which may change faster than government requirements.

The discussion also revealed a sense that industry has latched too much onto Software Bill of Materials (SBOMs) (see Section~\ref{sec:sbom}) and not enough onto other aspects, including attestations (see Section~\ref{sec:attestation}).
For example, the EO requires attestation that companies follow the NIST Secure Software Development Framework (SSDF)~\cite{ssdf}.
The attestation process is currently very manual; however, there is potential to automate many aspects of it.
One participant noted that companies are starting to automate different aspects of the attestation requirements and that the first one to do so will define the landscape.

Finally, there were comments and concerns about the fact that the EO only states that SBOMs should be created and does not describe what should be done with that data.
On the one hand,  a participant stated that the EO lacks a clearly articulated problem statement.
SBOMs have existed. Specifications have existed. There is an incredible amount of literature that we are not using in the workshops.
On the other hand, there was the recognition that each system is different.
A risk for one might be different than a risk for another.
This makes it complicated to prescribe specific SBOM uses.
One participant noted that a key value of the EO is that we are speaking openly that we are operating in an adversarial environment.

\subsection{Open Questions}
At the end of the panel, some open questions remained:

\begin{itemize}
\item How can attestations be made more automated and scalable? Companies are starting to do this already. The first there will define the landscape.
\end{itemize}

\section{Software Bill of Materials (SBOM)}
\label{sec:sbom}

An SBOM is a nested inventory of 'ingredients' that make up the software component or product that helps to identify and keep track of third-party components of a software system. The EO states that any company that sells software to the federal government must issue a complete SBOM that complies with the National Telecommunications and Information Administration (NTIA) Minimal Elements \cite{NtiA}.  

Vulnerability Exploitability eXchange (VEX) \cite{VEX} information was touched upon throughout the discussion.  An SBOM can be accompanied by a VEX \cite{VEX} addendum, which is a form of a security advisory that indicates whether a
product or products are affected by a known vulnerability or vulnerabilities. 

\subsection{Current State of the Practice}
Increasingly, tool vendors and package managers are providing the capability to generate SBOMs. Participants acknowledged that SBOM output is not consistent between tools. Additionally, tool support is lagging for embedded software. Participants expressed a desire for SBOMs to be signed, which is not currently the state of the practice.

Tools to consume and provide actionable information from SBOMs are lagging tools to produce SBOM.  As a result, participants acknowledged it could take years to use SBOMs effectively to aid in response and recovery from a cyber event.  Some expressed concern that information in SBOMs may make it easier for adversaries.

\subsection{Open Questions}
At the end of the panel, some open questions remained:

\begin{itemize}
\item What are the use cases for SBOM?  What is the benefit of benefit gained by having an SBOM?
\item How will SBOMs be shared, including how will they be shared across agencies so companies do not have to share separately with each agency they do business with?
\item What other kinds of Bills of Materials should be considered, such as Firmware Bill of Materials (FBOM) or Hardware Bill of Materials (HBOM)?
\end{itemize}

\section{Choosing Dependencies}
\label{sec:deps}

Open source dependencies vary widely in quality, maintenance, origin, and licenses.
Every dependency introduces value and risk, and 
once a dependency is incorporated into a project, it is often hard to replace.
Therefore, it is important to have a policy that governs how software developers may choose new dependencies.

\subsection{Consumers of Software}

``Dependencies'' can be thought of at different levels.
In contrast to the industry summits, participants at the government summit are more often on the consumer side of software production.
They do not focus much on choosing or updating specific software library dependencies.
Their contractors do this.
That said, participants felt that the increased attention to the software supply chain has led to a significant increase in awareness of how much open source software they use.
This awareness leads to a better understanding of what they need to manage.

Having policies around choosing dependencies is also vastly different in the government space.
In contrast to an industry policy, a policy in a government organization comes with its own bureaucracy, and bureaucracy can stifle innovation.
Therefore, there was a significant hesitance to a permission-based culture around software development.
One participant reflected on a government policy decades ago that only allowed software development using the Ada programming language.

Finally, the participants discussed what it meant to use software dependencies from foreign countries, i.e., anything not developed in the United States.
There was discussion about efforts to identify and manage when software comes from embargoed countries.
However, there was also an awareness that adversaries can find ways around those mechanisms.
Overall, there was the sentiment that ``foreign'' doesn't mean bad; it means risk.

\subsection{Open Questions}

\begin{itemize}
    \item How can we be objective about the risk of dependencies with foreign origins?
\end{itemize}

\section{Self-Attestation and Provenance}
\label{sec:attestation}

As discussed in Section~\ref{sec:eo}, Executive Order 14028 requires government contractors to attest to 
(1) conformity with secure software development practices as well 
(2) the integrity and provenance of open-source software used within any portion of a product.
In the context of the software supply chain, provenance refers to not only the identity that created each dependency and transitive dependency but also the process through which each software component was built.
For example, provenance is a key part of the SLSA framework~\cite{SLSA}, and systems such as in-toto\footnote{\url{https://in-toto.io/}} can be used to capture and communicate provenance information.

\subsection{Attestation}

Earlier summit discussion (see Section~\ref{sec:eo}) touched upon industry's concerns about ambiguity in self-attestation requirements.
Contractors are asked to self-attest conformance to NIST's Secure Software Development Framework (SSDF)~\cite{ssdf}.
However, one participant noted that the SSDF is ``fuzzy buckets.''
There is not a lot of specificity and many degrees of freedom.
What it means to attest to the SSDF is a wide range, which is why industry is looking for something more concrete.

Participants also compared the SSDF conformance requirements to the (arguably) failed rollout of the original Cybersecurity Maturity Model Certification (CMMC) requirements.
Ultimately, self-attestation was not working for the DoD, so CMMC started requiring external parties to come in.
However, participants noted that SSDF is different than CMMC, because industry is already working towards machine-consumable ways of tracking the information that is part of the self-attestation.
Of specific note were efforts in the IETF and the OpenSSF's supply chain integrity group.
Many participants agreed that the SSDF and attestation are far more foundational to security than the SBOM. Similar to the discussion of SBOM product and consumption (see Section~\ref{sec:sbom}), participants acknowledged that processes to consume and share attestation data were lagging further behind the production of attestation data.  

\subsection{Provenance}

The participants had mixed feelings on provenance in the software supply chain.
For example, one participant noted that some partners really like to know who is contributing to projects.
However, others were concerned that policies based on provenance information simply are not enforceable.
There was also recognition that sometimes critically needed software might have origins buried within them that violate policies.
This is just the nature of open-source software.

Some participants also raised the sentiment that it is more important to verify quality than consider provenance.
As an anecdote, a participant noted that we should not have cared about Kaspersky from a provenance perspective.
Instead, the real concerns emerged when you considered the \emph{functionality} it was performing.
Another participant recalled incidents that were identified not because of provenance information, but because of a whistle-blower.

Finally, there were comparisons to provenance in hardware supply chains.
There was a discussion of supply chain illumination tools with concerns of false positive rates resulting from their probabilistic risk assessment approaches.
Given that many of these tools consider contractual elements, they apply more to closed-source software than open-source.

\subsection{Open Questions}
At the end of the panel, some open questions remained:

\begin{itemize}
\item To what extent can the evaluation of conformance to the SSDF be automated?
\item To what extent should we focus on provenance versus other indicators of security?
\item How will attestation data be shared and consumed?
\end{itemize}

\section{Large Language Models (LLMs)}
\label{sec:llm}

Within the last year, Large Language Model (LLM)-based systems, such as ChatGPT, are increasingly being used for automated code generation.   

\subsection{Security Concerns}
The popularity of LLM-based systems for code generation causes a security concern of adversarial models resulting in tainting the training data such that vulnerable code is generated and infiltrated into software systems.  
A participant expressed concerns that the public's accelerated use of LLM can lead to large-scale data exfiltration.  LLM system users regularly pull LLM output into their product and regularly contribute their own proprietary data into training data through their queries. 

Participants also expressed concern about the use of social engineering of LLM-based tools.  For example, a participant knew of a country that input a policy into an LLM system for language translation. Now that country's official policy may be considered to be generated by a computer and may have subtle unintended differences when compared with the original that could cause unrest. There is concern about the use of LLMS to manipulate public opinion and to manipulate society. Some government agencies do not allow LLM use on government computers.   

\subsection{Open Questions}
At the end of the panel, some open questions remained:

\begin{itemize}
\item Is it possible to quantify the vulnerabilities in LLM training data?
\item How can LLMs be used to aid in software supply chain security?  Can LLMS enable whole proof generation and proof repair?
\end{itemize}

\section{Misc Discussion}

Throughout the Summit, several additional topics were discussed, primarily originating from the sharing of the two prior industry summits.

\subsection{Updating Dependencies}
Most software uses a plethora of third-party dependencies that provide common functionality and help developers with productivity. However, these dependencies add an additional layer of complexity and lead to an ecosystem of direct and transitive dependencies that each software replies on. A security vulnerability in a third-party dependency can lead to cascading issues and needs to be updated with the newly released version fix as soon as possible. 

As primarily software consumers, the government participants acknowledge the complexities of software suppliers needing to update dependencies. They expressed being even more nervous about updating vulnerable containers than with software components due to the processes of scanning and updating containers being less mature than components.

 \subsection{Secure Build and Deploy}
 Various build platforms and CI/CD tools support developers in automating the parts of software development related to building, testing, and deploying. These platforms further help in enhancing software build integrity by establishing documented and consistent build environments, isolating build processes, and generating verifiable provenance.  Participants shared that agencies may have a process for evaluating and ``blessing'' tools.  The agencies may enforce that vendors use approved tools in their CI/CD pipeline.  
\section{Executive Summary}
Though some government agencies have large software development teams, most are primarily consumers of software and, therefore, consumers of the software development practices and artifacts mandated by the EO. The participants acknowledged that industrial organizations may be confused by the lack of specificity in the EO, including guidance on SBOM sharing and attestation production.   Work is underway to produce the specificity, but the lack of guidance may also provide implementation flexibility by organizations.  For example, a company that solves automating attestation information may be influential in establishing this specificity. Participants acknowledged the complexities organizations face in choosing and updating dependencies and provided a realistic perspective that an adversary in an embargoed country can find the means to disguise the geographic origins of the dependency.  Finally, LLMs are considered both a security risk and offer the potential to aid in securing the software supply chain.

\section{Acknowledgements}
A big thank you to all Summit participants. We are very grateful for being able to hear about your valuable experiences and suggestions. The Summit was organized by and recorded by Laurie Williams and William Enck.  This material is based upon work supported by the National Science Foundation Grant Nos. 2207008, 2206859, 2206865, and 2206921.
These grants support the Secure Software Supply Chain Summit (S3C2) consisting of researchers at North Carolina State University, Carnegie Mellon University, University of Maryland, and George Washington University. Any opinions expressed in this material are those of the author(s) and do not necessarily reflect the views of the National Science Foundation.
%\end{acks}

\bibliographystyle{ACM-Reference-Format}
\bibliography{literature}

\appendix

\section{Full Survey Questions for Panel}
\label{questions}
\begin{enumerate}
\item What effect has the Executive Order had on the government agencies, contracted and third-party suppliers you interact with?
\item From your perspective, how wide-spread is the practice of producing SBOMs? What will/can SBOMs actually achieve? How can they be leveraged/used?  Do you see VEX being helpful or hurtful?  In what ways?
\item From the perspective of the government agencies you interact with, what policies (if any) do you see around choosing new dependencies and bringing them into the project and developer workstations? Has this changed in the last few years? If so, how? Are you seeing adoption of OpenSSF Scorecard or other metrics to help you with your decision making?
\item From the perspective of government agencies you interact with, to what extent is provenance information being produced and/or consumed? From the perspective of government agencies you interact with, to what extent is self-attestation information being produced and/or consumed? 
\item From the perspective of government agencies you interact with, what is your perspective on the use of large language models (LLMs) such as ChatGPT as another supply chain attack vector?
\end{enumerate}

\end{document}